\newcommand{\PRL}[3]{Phys.\ Rev.\ Lett.\ {\bf #1},\ #2 (#3)}

\newcommand{\PRA}[3]{Phys.\ Rev.\ A\ {\bf #1},\ #2 (#3)}

\newcommand{\JPB}[3]{J.\ Phys.\ B:\ At.\ Mol.\ Opt.\ Phys.\ {\bf #1},\ #2 (#3)}

\documentclass[aps,showkeys,showpacs,amsmath,amssymb]{revtex4}
\usepackage{dcolumn}
\usepackage{longtable}
\begin{document}
\title{Minimal classical communication and measurement complexity for quantum information splitting of a two qubit state}
\pacs{03.67.Hk, 03.65.Ud}
\keywords{Entanglement, Teleportation, State sharing, Superdense coding}
\author{Siddharth Karumanchi}
\affiliation{Birla Institute of Technology and Science, Pilani, Rajasthan- 333031, India}
\author{Sreraman Muralidharan}
\email{sreraman@loyolacollege.edu} \affiliation{Loyola College,
Nungambakkam, Chennai - 600 034, India}
\author{Prasanta K. Panigrahi}
\email{prasanta@prl.res.in}
\affiliation{Indian Institute of Science Education and Research (IISER) Kolkata, Salt Lake, Kolkata - 700106, India}
\affiliation{Physical Research
Laboratory, Navrangpura, Ahmedabad - 380 009, India}
\begin{abstract}
We provide explicit schemes for quantum information splitting (QIS) of real or equatorial two qubit states
among two parties, through the five particle cluster and Brown states respectively. The schemes introduced
are neither fully deterministic nor completely probabilistic and are carried out in the case
where, the information
that is to be shared is partially or fully known to the sender. We show that, QIS can be accomplished with just two classical bits,
as against four classical bits, when the information is unknown to the sender. 
\end{abstract}
\maketitle
\section{Introduction}
Entanglement is the most striking and counter intuitive feature of quantum mechanics that has come in handy
for a number of practical applications in communication technology \cite{Nielson}. It is well understood only in the case 
of two particles. In the multiparticle scenario, much remains to be understood and explored owing to the
increase in complexity with the number of qubits as there are numerous ways in which they can be
entangled. Entangled states are used as a communication resource in teleportation, secret sharing
and superdense coding. Quantum teleportation is a technique for transfer of information
between parties, using a distributed entangled state and a
classical communication channel. In a remarkable work, Bennett \it et al. \normalfont \cite{Bennett} showed that, an EPR pair,
could be used for the teleportation of an unknown single qubit state $\alpha|0\rangle+\beta|1\rangle$, where
$|\alpha|^2 + |\beta|^2 = 1$. Initially, Alice the sender, combines the unknown qubit state
, with the EPR pair and performs a Bell measurement, and conveys the outcome of her measurement to Bob the receiver,
via two classical bits of information. Bob, then performs appropriate unitary transformations on his particles and 
obtains the unknown qubit information. Hence, the resources required to achieve this are one
ebit of entanglement and two classical bits of information. Recently, attention has turned 
towards the teleportation of an arbitrary two qubit state \cite{Rigolin1, Yeo, Sreraman, Sreraman2} given by :  
\begin{equation}
|\psi\rangle=\alpha|00\rangle+\gamma|10\rangle+\beta|01\rangle+\delta|11\rangle,
\end{equation}
  where , $|\alpha|^2+|\beta|^2+|\gamma|^2+|\mu|^2 = 1$. The resource used for teleporting this are two ebits and four classical bits. 
  
  Quantum information splitting or secret sharing is the splitting up of quantum information between
various parties, such that if only each of them cooperate, one of them can obtain the desired information.
Hillery \it et al. \normalfont \cite{Hillery}, have demonstrated the sharing of an unknown single
qubit of information among three parties using the three and the four particle GHZ states.
The scheme uses, two cbits of information resources. Recently, QIS of an arbitrary two qubit state
was proposed by two of the present authors, utilizing the five particle Brown and cluster states \cite{Sreraman, Sreraman3}. 
The procedure utilizes
four cbits as information resource.
In all these schemes, the state that is to be shared is unknown to the sender. However, 
in many practical scenarios, the information that is to be shared may be known to the sender.
In such cases, the classical information resource can be greatly reduced and need not be wasted. 
Recently, it was shown that if the sender knows the single qubit information that she is sending, it could be split 
among two parties by utilizing just one classical bit \cite{Zhang}. In this paper, we extend these 
results to an arbitrary two qubit state by investigating the five particle cluster and the Brown state
as entangled resources. Let Alice, possess the two qubit information, in Eq.(1) which she wants Bob and 
Charlie to share and be aware of the information. Here, 
$\alpha$ is real, while the other coefficients are complex. If we choose, $\alpha = \gamma = \frac{1}{2}$, 
$\beta = \delta = \frac{1}{2} e^{i\phi}$, then $|\psi\rangle$ is analogous to the equatorial state on the Bloch sphere.
It is to be noted that, if $\phi = 0$ or $\pi$,  then all the coefficients are real, in which case $|\psi\rangle$ is called a real state.

\section{Cluster state for QIS of a two qubit state}
The five particle cluster state is given by,
\begin{equation}
|C_5\rangle = \frac{1}{2} (|00000\rangle + |00111\rangle + |11101\rangle + |11010\rangle).
\end{equation}
There are two ebits of entanglement between pairs $15|234$. Hence, we let Alice posses qubits, 1 and 5 and Bob possess
the remaining qubits. Since, Alice is aware of the information that is to be sent, she can perform a two particle measurement in the 
basis involving the coefficients of the input state. The outcome of the measurement performed by Alice and 
the state obtained by Bob and Charlie are shown in the table below 

 \begin{table}[h]
\caption{\label{tab1} The outcome of the measurement performed by Alice and the state obtained by Bob and Charlie}
\begin{tabular}{|c|c|}
\hline {\bf Outcome of the measurement } & {\bf State obtained }\\
$\frac{1}{2}(\alpha|00\rangle + \beta|01\rangle + \gamma|10\rangle + \delta|11\rangle)$&$\alpha|000\rangle + \beta^*|011\rangle + \gamma|101\rangle + \delta^*|110\rangle$\\
$\frac{1}{2}(\beta^*|00\rangle - \alpha|01\rangle + \delta^*|10\rangle - \gamma|11\rangle)$&$ \beta|000\rangle - \alpha|011\rangle + \delta|101\rangle - \gamma|110\rangle$\\
$\frac{1}{2}(\gamma|00\rangle - \delta^*|01\rangle - \alpha|10\rangle + \beta^*|11\rangle)$&$\gamma|000\rangle - \delta|011\rangle - \alpha|101\rangle + \beta|110\rangle$\\
$\frac{1}{2}(\delta^*|00\rangle + \gamma|01\rangle - \beta^*|10\rangle - \alpha|11\rangle)$&$\delta|000\rangle + \gamma|011\rangle - \beta|101\rangle - \alpha|110\rangle$\\
\hline
\end{tabular}
\end{table}	
Neither Bob nor Charlie can obtain the information through local operations on their own qubits.
Alice encodes the outcome of her measurement in classical bits and sends it to Charlie. Bob can now perform
a single particle measurement and send the outcome of her measurement to Charlie. Each of the measurement
can be performed with equal probability ($=\frac{1}{4}$). If Alice performs a measurement in the first basis,
it is not possible for Charlie to obtain the desired state, unless the information to be shared is equatorial
 or real qubit states. If Alice chooses to perform a measurement in remaining members of the basis, then Charlie can 
 obtain the desired state after receiving Bob's measurement output as in the standard QIS scheme. In this sense,
 the present algorithm is neither fully deterministic nor probabilistic. 
  Now, Bob can perform a measurement in the basis $\frac{1}{\sqrt{2}} (|0\rangle+|1\rangle$,
 then Charlie's system collapses to $\frac{1}{2}(00\rangle + e^{-{i\phi}} |11\rangle \pm |01\rangle \pm e^{-{i\phi}} |10\rangle)$, 
To correct the phase factor, Charlie has to use the gate controlled phase of $e^{(2i\phi)}$
i.e., when the first qubit is $|1\rangle$ then a phase of $e^{{2i\phi}}$ is introduced.

In the original scheme, using cluster state for QIS of an arbitrary two qubit state,
Alice performs a four particle measurement and conveys the outcome of her measurement to 
Charlie via four cbits of information. In comparison, the present scheme uses only two cbits of information, if the
two qubit state is chosen from partially known equatorial or real states.  Hence for these special states
 this procedure reduces measurement complexity and the consumption of classical information by two cbits.

\section{Brown state for QIS of a two qubit state}
The Brown state is given by 
\begin{equation}
|\psi_{5}\rangle=\frac{1}{2}(|001\rangle|\phi_{-}\rangle+|010\rangle|\psi_{-}\rangle
+|100\rangle|\phi_{+}\rangle+|111\rangle|\psi_{+}\rangle),
\end{equation}
where, $|\psi_\pm\rangle=\frac{1}{\sqrt{2}}(|00\rangle\pm|11\rangle)$ and
$|\phi_\pm\rangle=\frac{1}{\sqrt{2}}(|01\rangle\pm|10\rangle)$ are Bell states.
In this scheme, we let Alice have first two particles, Bob have the third one and
Charlie the last two particles. As in the previous case, in this scheme also
Alice can perform a measurement in appropriate basis. The outcome of the measurement
performed by Alice and the Bob-Charlie system is shown below.

 \begin{table}[h]
\caption{\label{tab1} The outcome of the measurement performed by Alice and the state obtained by Bob and Charlie}
\begin{tabular}{|c|c|}
\hline {\bf Outcome of the measurement } & {\bf State obtained }\\
$\frac{1}{2}(\alpha|00\rangle + \beta|01\rangle + \gamma|10\rangle + \delta|11\rangle)$&$\alpha|\eta_1\rangle + \beta^* |\eta_2\rangle
+ \gamma |\eta_3\rangle + \delta^*|\eta_4\rangle$\\
$\frac{1}{2}(\beta^*|00\rangle - \alpha|01\rangle + \delta^*|10\rangle - \gamma|11\rangle)$&$\beta|\eta_1\rangle + \alpha |\eta_2\rangle
+ \delta |\eta_3\rangle - \gamma|\eta_4\rangle$\\
$\frac{1}{2}(\gamma|00\rangle - \delta^*|01\rangle - \alpha|10\rangle + \beta^*|11\rangle)$&$\gamma|\eta_1\rangle - \delta |\eta_2\rangle
- \alpha |\eta_3\rangle + \beta|\eta_4\rangle$\\
$\frac{1}{2}(\delta^*|00\rangle + \gamma|01\rangle - \beta^*|10\rangle - \alpha|11\rangle)$&$\delta|\eta_1\rangle + \gamma |\eta_2\rangle
- \beta|\eta_3\rangle - \alpha|\eta_4\rangle$\\
\hline
\end{tabular}
\end{table}	
Here,
\begin{eqnarray}
|\eta_1\rangle = \frac{1}{{2}}(|101\rangle-|110\rangle),\\
|\eta_2\rangle = \frac{1}{{2}}(|000\rangle-|011\rangle),\\
|\eta_3\rangle = \frac{1}{{2}}(|001\rangle+|010\rangle),\\
|\eta_4\rangle = \frac{1}{{2}}(|100\rangle+|111\rangle).\\
\end{eqnarray}

Alice can encode the outcome of her measurement in two cbits and send it to Charlie. As in the previous case, even here, if 
the first member of the set is chosen as the measurement basis, then one can obtain the
desired information, only if the initial state is either equatorial or real. Bob then performs a 
measurement in the basis $\frac{1}{2} (|0\rangle \pm |1\rangle)$. The corresponding states
obtained by Charlie are  : 
\begin{equation}
\pm (01\rangle - |10\rangle) + e^{-i\phi} (|00\rangle-|11\rangle) + (|01\rangle+|10\rangle) \pm e^{-i\phi} (|00\rangle + |11\rangle)], 
\end{equation}
on which Charlie applies a suitable unitary operator to obtain the state. Hence, the protocol suceeds.

\section{Conclusion}

In conclusion, we have shown that, for the QIS of a two qubit state, the classical information resource
is reduced by half, if the two qubit state is equatorial or real. Hence, one need not waste 
classical information resource, when some information is known about the initial state. Interestingly, the scheme is
neither fully probabilistic nor deterministic. It is worth checking, if classical
or quantum information resources in other protocols can be reduced when some information
about the initial state is known to the sender. Generalization to the QIS of 
real or equatorial $N$ qubit states using $2N$ and a $(2N+1)$ qubit states as an entangled 
resource needs to be explored.

\end{document}